# Point-Cloud Based Inverse Design of Free-Form Metamaterials Using Deep Generative Networks

Kijung Kim[1†], Seungwook Hong[1†], Wonjun Jung[1], Wooseok Kim[1], Namjung Kim[2*], and Howon Lee[1*]

[1]Department of Mechanical Engineering, Institute of Advanced Machines and Design, Seoul National University, Seoul, Republic of Korea
[2]Department of Mechanical Engineering, Gachon University, Sungnam, Republic of Korea

[*]To whom correspondence should be addressed: namjungk@gachon.ac.kr, howon.lee@snu.ac.kr
[†]These authors contributed equally to this work.

**Abstract**

Mechanical metamaterials enable precise control over structural properties, but their design method remains challenging due to their complex structure. Although additive manufacturing has expanded geometric freedom, navigating this vast and complex design space still requires computationally intensive simulations or expert-driven processes. Recently, artificial intelligence (AI)-driven design approaches have emerged to address these limitations, but many studies restrict their scope to parametric representations, limiting their generative capacity to predefined shapes. Here, we present a point cloud-based generative framework that enables the inverse design of 3D metamaterial without parametric constraints. Trained on a number of structurally valid unit cells, the present machine learning model learns geometric patterns, mitigates common connectivity issues inherent in point cloud generation. The proposed model constructs a latent space organized by mechanical properties and naturally clustered by unit cell types. By sampling this latent space, our method supports both property-guided inverse design and generation of topologically gradient transition between distinct unit cell types. This approach facilitates inverse design of 3D metamaterials with high geometric complexity.

**Introduction**

Mechanical metamaterials are artificially engineered materials which exhibit unconventional mechanical behaviors not found in nature. Unlike conventional materials, the mechanical characteristics of metamaterial arise not by material composition but from the geometry of their internal structure, which is often realized through lattice structures composed of periodically arranged unit cells. This geometry-driven tunability enables the precise tuning of a wide range of mechanical properties, including high strength-to-weight performance (*1, 2*), negative Poisson's ratios (*3, 4*), and negative coefficients of thermal expansion (*5*). Despite this potential, the fabrication of complex microlattice structures has been constrained by the limitations of traditional manufacturing, such as insufficient resolution and the inability to produce fine features. Recent advancements in additive manufacturing (AM) significantly alleviated these constraints, allowing the precise fabrication of complex microstructures across various size scales (*1, 6*). This capability of AM has enabled the fabrication of a diverse metamaterial structures beyond conventional limitations. Consequently, there has been growing interest in the design of novel metamaterial structures with highly customized mechanical properties for diverse applications (*7*).

Despite advances in fabrication, the design of microlattice structures remains challenging. Traditional design approaches include mechanics-based analysis, topology optimization, and bioinspired designs. Mechanics-based analysis relies on classical mechanical theories and is typically limited to simple structures due to analytical intractability (*8, 9*). Topology optimization offers greater flexibility (*10, 11*), but requires significant computation resources due to iterative evaluations. Bioinspired approaches imitate the designs found in nature (*12*), but translating these into manufacturable and robust structures is often challenging. These methods generally demand extensive prior knowledge, rely on computationally expensive simulation, or involve substantial trial and error refinement. In response, machine learning (ML) has emerged as a promising paradigm for metamaterial design, offering data-driven alternatives that can learn complex

structure–property relationships directly from data (*13-24*). Building on this capability, ML-based methods are able to efficiently explore high dimensional design spaces, and enabling inverse design, which generates structures from predefined performance targets. Several studies have demonstrated the advantages of various ML in metamaterial design, including neural networks (*15*), Generative Adversarial Networks (GANs) (*17, 19*), and ML-assisted topology optimizations (*22*). However, these methods remain limited in their ability to handle complex 3D structures, largely due to difficulty of representing intricate geometries in a format suitable for learning and generation. Many of existing studies rely on 2D geometries or simplified 3D structures to reduce computational cost and ensure training stability. However, such simplifications inherently restrict the design space and fail to capture irregular or highly detailed topologies.

A key challenge in applying ML to 3D metamaterial design lies in how geometric structures are represented. Three commonly used formats to represent 3D structures - voxel, graph, and point cloud - offer different advantages and limitations. Voxel representation distinguish solid and void regions based on 3D space of uniformly divided spatial grids and provides compatibility with 3D convolutional neural networks (*25, 26*). However, they suffer from staircase artifacts when representing curved geometries, and require high computational cost to achieve fine resolution. Graph representation encode structural connectivity through nodes and edges, offering compactness and interpretability (*27-30*). But they typically rely on simplified assumptions such as uniform thickness or predefined node positions. In contrast, point cloud represent 3D structures as sets of spatially distributed points, enabling high-resolution encoding of complex, nonparametric geometries (*31-35*). Unlike voxel representation which encode both solid and void regions, point clouds store only the solid geometric information, significantly reducing computational burden. These advantages make point clouds a suitable candidate for generative design frameworks capable of capturing intricate geometries at high resolution. However, their adoption in 3D metamaterial design has remained limited due to inherent challenges, such as lack of connectivity information between points, and incompatibility with standard neural network architectures, which can compromise structural validity and learning efficiency.

In this study, we present a point cloud based deep generative network for the design of mechanical metamaterials with targeted mechanical properties. As illustrated in **Figure 1**, we construct a diverse point cloud dataset of structurally valid unit cells, which are derived from widely known 3D lattice types, each paired with mechanical property obtained via finite element analysis. Generative model based on a variational autoencoder (VAE) is trained to embed these geometries into a continuous latent space, where data points are naturally clustered by topology and organized by mechanical performance. This framework enables property-guided inverse design through latent space sampling. To address the connectivity challenge of point cloud, the model implicitly learns structural validity by referencing a large number of valid structural samples during training. Furthermore, latent interpolation between clusters allows for the generation of novel unit cell geometries with gradually changing topology, leading to topologically graded structures showing smooth transitions between different unit cell types. The generated point cloud designs were fabricated by 3D printing and experimentally validated through mechanical testing. The results confirmed that their mechanical properties closely matched the target value. This study establishes a novel point cloud-based generative framework for 3D metamaterials, enabling both inverse property-guided design and the creation of topologically graded structures. Our method offers a scalable and data-driven alternative to conventional design approaches.

## Results

**Training dataset: point cloud representation of unit cells and mechanical property**

Point clouds provide a nonparameterized and high-resolution representation for complex metamaterial geometries. However, the absence of explicit connectivity information between points is a major limitation that can compromise the structural integrity of generated designs. Since ML models operate in a data-driven manner, their ability to generate physically valid structures strongly depends on the structural consistency of the training data. Therefore, it is crucial to construct a dataset that preserves both geometric connectivity and sufficient diversity for learning metamaterial topologies. To address this, we constructed a point cloud dataset based on six existing unit-cell topologies that are widely used in mechanical metamaterial design: body centered cubic (BCC), cubic, octahedron, octet-truss, Kelvin foam, and fluorite structure, as shown in **Figure 2(A)**. These well-established structures inherently satisfy connectivity constraints, enabling the model to implicitly learn valid geometric relationships through training.

To generate diverse structural variations from each unit-cell type, we parametrized the geometry with design variables. **Figure 2(B)** illustrates the data generation process of the BCC structure as an example. A BCC unit cell, composed of four diagonal struts, is defined by three data parameters: *a*, *b*, and *r*. Parameters *a* and *b* represent the width and height of the unit cell, respectively, while the length is fixed at 30 (arbitrary unit). The parameter *r* corresponds to the strut radius. By assigning specific values to parameters *a*, *b*, and *r*, solid geometries of the BCC structure can be generated as shown in the first image of **Figure 2(B)**. A voxelization process was then applied in the constructed geometries, to define the spatial points constituting the structure. The geometry was embedded into an evenly spaced 3D voxel grid, and each voxel center was evaluated to determine its distance from structure. Points inside the solid region were retained to form the point cloud representation of the BCC structure. To reduce data redundancy, we leverage the symmetry of the unit cell and excluded mirrored regions through a preprocessing step. This exclusion effectively compresses the point cloud to one-eighth of its original size, as shown in the final image of **Figure 2(B)**. This process translates a given set of parameters *a*, *b* and *r* to a point cloud data representing a corresponding unit cell geometry. To construct a diverse dataset based on this process, we randomly varied *a* and *b* within the range of 16 to 60, which corresponds to half to twice the fixed length of 30. The strut radius *r* was varied from 1 to 3. In this way, we systematically generated 10,000 structurally valid and diverse BCC structure samples. The same procedure was applied to all six unit-cell types, resulting in a consistent and diverse dataset for training. This dataset preserves structural connectivity of the original unit cell geometries while capturing geometric diversity across unit-cell classes. Although we impose certain restrictions on the design parameters for practical considerations, the framework can be extended to accommodate diverse application-specific requirements. Further details about the parameters of other unit cell types are provided in the **Figure S1** in **Supplementary Materials**.

To obtain mechanical property labels for each generated point cloud geometries, we calculated the effective stiffness of each unit cell using homogenization-based finite element analysis (FEA) (*36*). The FEA code used in this work takes voxel data of the geometry and material properties as input, and computes a 6×6 stiffness matrix (***C***) that characterizes the effective elasticity tensor of the unit cell. Since the point cloud structures were originally derived from voxel geometries, we used the voxel data as input for this FEA process, as shown in **Figure 2(C)**. The material properties, such as Young's modulus and Poisson's ratio, were set to 2.8 GPa and 0.3, respectively, to match those of the 3D printing material used in experiments (described in detail later). The stiffness matrix obtained from the FEA provides various effective mechanical properties of the unit cell. Among these, we selected the Young's modulus in the z-direction ($E_z$), as the target property for training, which serves as a representative indicator of uniaxial compressive stiffness.

$$E_z = \frac{1}{S_{33}}, \quad \text{where } S = C^{-1} \tag{1}$$

The value of $E_z$ was extracted from the stiffness matrix as shown in **Equation (1)**, where $S_{ij}$ is a (*i,j*) component of the compliance matrix (***S***) derived from the stiffness matrix, ***C***. Other stiffness components, such as shear modulus could also be used depending on the target property expected from the unit cell to be designed. Through this process, modulus $E_z$ value of each generated point cloud geometry can be extracted. These geometry-property pairs then construct a labeled dataset for supervised training of the generative model. This dataset construction pipeline was applied to all six unit-cell types to create a comprehensive library that enables data-driven design of mechanically functional structures.

**Deep generative models**

Generative ML models provide a data-driven approach for exploring the complex design space of constructed point cloud datasets. Among various frameworks, we selected a variational autoencoder (VAE) (*23*) as a base model, due to its ability to learn a continuous and interpretable latent space (*37*). While alternatives such as GANs can also generate high-quality outputs, they do not provide a well-defined latent space, making inverse design and precise control over latent representations challenging. Moreover, point cloud data, which is unordered and lacks connectivity information, poses additional challenges for discriminator model in GANs, which generally rely on permutation-sensitive network architectures. In contrast, the VAE is able to capture essential geometric features of the point cloud, encodes them into a latent vector, and decodes them back to their original geometry. The overall model architecture is illustrated in **Figure 3(A)**, where each pair of point cloud geometry ($x$) and its property label are encoded into a latent vector ($z$), which is fed into the decoder for reconstructing the point cloud ($\hat{x}$). Then, reconstructed point cloud is mirrored to form the final unit-cell structure. Latent space representation of model provides a compact representation of point cloud data, which supports geometry-aware and property-guided sampling, laying the foundation for integrating property prediction directly into the generative process.

To guide the latent space into property-aware manner, we incorporated a property regressor that maps the latent vector ($z$) to its corresponding stiffness property ($E_z$) (*38*). Trained as a surrogate model, the regressor predicts the mechanical stiffness of a given structure directly from its latent representation, offering a computationally efficient and fast alternative to the FEA simulation commonly used in traditional design workflows. To validate the model, we present the results obtained from the dataset of BCC structures, as a representative case. **Figure 3(B)** shows the performance of the regressor model, where the predicted property values on the y-axis closely match the actual property values from the training data on the x-axis, confirming the strong predictive accuracy. This regressor was trained jointly with the VAE, to encourage structures with similar properties to cluster together in the latent space, thereby achieving property-aware alignment of latent space. **Figure 3(C)** illustrates a two-dimensional principal component analysis (PCA) visualization of the latent vectors acquired from VAE-regressor joint model, trained on the dataset of BCC structures. Each point represents a single unit cell geometry from the training dataset, and the point color indicates the mechanical property of the corresponding unit cell. The distribution shows that unit cells with similar properties are positioned closer together in the latent space, demonstrating the property-aware alignment achieved through joint training of the regressor. This alignment sets the basis of the proposed inverse design of metamaterials, in which latent vectors from regions corresponding to the desired property can be sampled in the latent space and reconstructed to the associated unit cell structures through the decoder.

**Loss functions**

The VAE and regressor were trained using four loss components: reconstruction loss ($\mathcal{L}_{reconsturciton}$), Kullback–Leibler divergence for regularization loss ($\mathcal{L}_{regularization}$), regressor loss ($\mathcal{L}_{regression}$), and contrastive loss ($\mathcal{L}_{contrastive}$), weighted by $1, \beta, \gamma$ and $\delta$, respectively, as described in **Equation (2)**.

$$\mathcal{L} = \mathcal{L}_{reconsturciton} + \beta \times \mathcal{L}_{regularization} + \gamma \times \mathcal{L}_{regression} + \delta \times \mathcal{L}_{contrastive} \qquad (2)$$

The reconstruction loss and regularization loss constitute the basic loss terms of the VAE. In the VAE framework, the input is compressed into a latent vector through the encoder and then reconstructed through the decoder, with the reconstruction loss defined as the difference between the input and its reconstruction. Since our study employs point cloud data, the chamfer distance, $d_{CD}$, which assesses how closely the reconstructed point clouds match the original inputs, is used as the reconstruction loss. The equation for this loss term is defined in **Equation 3**, sums the nearest point distances between the two sets of points ($S_1$, $S_2$), ensuring that the reconstruction quality of VAE maintains the spatial integrity and distribution of the original data (*37*).

$$d_{CD}(S_1, S_2) = \sum_{x \in S_1} \min_{x' \in S_2} \|x - x'\|_2^2 + \sum_{x \in S_2} \min_{x' \in S_1} \|x - x'\|_2^2 \tag{3}$$

Regularization loss is used to prevent overfitting of the training and ensures a smooth and continuous latent space, which is essential for meaningful interpolation and stable generation of new samples. Generally, KL divergence term is used to regularizes the learned latent distribution $q_\theta(x|z)$ to be close to a prior distribution $p(z)$, typically a standard normal distribution. Combining these two losses forms the basic VAE loss function, the evidence lower bound (ELBO), described in **Equation (4)**.

$$\mathcal{L}(\theta, \varphi) = -E_{z \sim q_\theta(Z|X)}\left[log\left(p_\varphi(x|z)\right)\right] + D_{KL}[(q_\theta(x|z)|p(z))] \tag{4}$$

The first term corresponds to the reconstruction loss, represented by the chamfer distance in this work, and the second term corresponds to the regularization loss, given by the KL divergence.

Next, to improve the accuracy of the regressor, the regression loss is calculated as the mean square error (MSE) between the actual properties of the dataset and the properties predicted by the regressor. Therefore, through the joint training of VAE and regressor, a continuous structured latent space is constructed where data with similar property are located near each other in the latent space.

Since our inverse design approach relies on sampling latent vectors corresponding to specific target values from an aligned latent space, making accurate and smooth property-aware alignment is essential. Therefore, a contrastive loss term is incorporated as an additional loss function into the training. The contrastive loss works by reducing the distance between similar data points while increasing the distance between dissimilar ones in the latent space. This method encourages data with similar properties to cluster closer together in the latent space, which complements the role of the regressor and further strengthen the alignment of the latent space. Exact equation for loss term is described in **Equation (5)**.

$$\mathcal{L}_{contrastive} = \frac{1}{n^2}\sum_{i=1}^{n}\sum_{j=1}^{n}[S_{ij} \cdot D_{ij}^2 + (1 - S_{ij}) \cdot \max(margin - D_{ij}, 0)^2] \quad (5)$$

where $n$ represents the total number of point cloud data points, $S_{ij}$ is the similarity matrix and $D_{ij}$ is the Euclidean distance between the latent vectors $z_i$ and $z_j$ (*39-41*). The similarity matrix is $S_{ij}$ defined such that $S_{ij} = 1$ if $|y_i - y_j| \leq$ threshold, indicating that $y_i$ and $y_j$, which represents the properties of the respective data, is similar. Otherwise, $S_{ij} = 0$ when the property difference between two data exceeds the threshold, indicates dissimilarity. Furthermore, the '$margin$' parameter determines the required minimum distance between dissimilar data within the latent space. By training the VAE with the contrastive loss, the model is optimized to position data with similar properties closer together in the latent space. This approach enables the VAE to learn not only the reconstruction of the data but also the relationships between properties and data, enhancing its overall learning capability. The results obtained with different weight parameters for each loss term, as well as the detailed parameter optimization process, are provided in **Figure S2** in **Supplementary Materials**.

**Inverse design of unit cells with targeted properties**

In this section, we demonstrate the inverse design process using proposed VAE-regressor framework and verify its effectiveness. To validate the approach, we first selected BCC as an example training dataset because of its geometric simplicity. During training, key hyperparameters, such as loss function weights and latent dimension were optimized to minimize the overall loss.

This optimization enabled accurate point cloud reconstruction by the VAE, precise property prediction by the regressor, and a well-aligned latent space with respect to the target property $E_z$ values. Detailed results, including the effect of different hyperparameter settings on loss reduction, latent space alignment, property prediction accuracy, and point cloud reconstruction, are provided in **Figure S2(F)** in **Supplementary Materials**. Based on this trained model with an optimized latent space, we then proceeded to generate new unit cell structures with specified $E_z$ values through an inverse design process.

To implement the inverse design, we generated new latent vectors by interpolating between pairs of existing latent vectors, which are expected to inherit interpolated property of their reference vectors. Therefore, the latent vectors interpolated to desired $E_z$ value are expected to generate structure that correspond to the specified mechanical properties. However, if two reference vectors are located far apart in latent space, reconstruction result of interpolated vector can be inaccurate or geometrically implausible. Therefore, rather than interpolating directly between the latent vectors with maximum and minimum $E_z$, we constructed a smooth transition path in latent space that captures gradual changes in property and geometry. This path was constructed by connecting latent vectors in increasing order of $E_z$, using a k-nearest neighbor graph method to ensure that neighboring reference points are located closely together and thereby maintaining geometric similarity. **Figure 4(A)** illustrates the resulting path, black lines with arrows represent the path from the start point (minimum $E_z$) to the end point (maximum $E_z$). Six colored circles indicate the selected reference train data for interpolation, which are distributed along the path to form a smooth transition.

For inverse design of target $E_z$ value, we selected a pair of adjacent reference points, whose predicted values are similar with the target. A new latent vector was then generated by linear interpolation between them. For example, when the target $E_z$ is 0.05 GPa, interpolation is performed on the path between two intermediate reference points with predicted values of 0.04 and 0.06 GPa, rather than between the global minimum and maximum. Due to similarity of reference points in both geometry and property, the generated latent vectors remain valid and structurally meaningful even though they do not exist in the original training dataset. In **Figure 4(A)**, white circles indicate newly generated latent vectors, each interpolated between a pair of reference train data. The generated latent vectors then passed through the regressor for validation and decoded into point cloud geometries, completing the inverse design workflow.

To validate the inverse design results, we compared the mechanical properties of the newly generated structures against target values. **Figure 4(B)** presents the predicted and computed $E_z$ values for ten generated samples, labeled $G_1$ through $G_{10}$, each corresponding to a predefined target value ranging from 0.01 to 0.10 GPa in increments of 0.01. For each sample, three types of $E_z$ values are shown: (1) *target property*, value used as the inverse design objective; (2) *ML prediction*, obtained by passing the interpolated latent vector through the regressor; and (3) *FEA*, calculated by applying FEA to the reconstructed point cloud geometry, obtained by decoder. As shown in the **Figure 4(B)**, ML predictions closely match the target values across all cases, indicating that the regressor successfully captures the latent–property relationship by training. The FEA results also exhibit strong agreement with both the ML prediction and target values. This alignment confirms the effectiveness of the model in generating structurally valid point clouds that yield the intended mechanical behavior. The high correlation between regressor predictions and FEA results (with an $R^2$ score of 0.99) further supports the reliability of the inverse design framework.

**Figure 4(C)** compares the design parameters of ten newly generated BCC unit cells ($G_1$–$G_{10}$, shown in orange) and six reference BCC unit cells (shown in blue), by arranging them from left to right in order of increasing $E_z$. Each unit cell geometry is shown with corresponding design parameters, *a*, *b* and *r*. Detailed values of each parameter and the corresponding $E_z$ are presented in **Table 1**. As $E_z$ increases, these parameters exhibit a smooth and continuous transition in shape, reflecting the effectiveness of interpolation process. Specifically, width (*a*) tend to decrease gradually, while the strut radius ($r$) increases. Remarkably, in the last two generated cases ($G_9$ and $G_{10}$), the strut radius exceeds the original dataset's upper bound of 3, indicating that the model can extrapolate beyond the training range to meet the targeted property. This result suggests that not only does the model interpolate geometry between reference structures, it also learns the function of key parameters to achieve the desired mechanical property.

To further validate the mechanical properties of the generated structures, we fabricated physical samples and performed compression tests. Four representative cases ($G_1$, $G_4$, $G_7$, and $G_{10}$) corresponding to target $E_z$ of 0.01, 0.04, 0.07, and 0.10 GPa, respectively, were selected for experimental validation. Using the design parameters extracted from the point cloud geometries shown in **Figure 4(C)**, 3D solid model of unit cell and 5×5×5 lattice structures for each case were constructed. **Figure 4(D)** presents these 3D models, and photographs of the fabricated samples via 3D printing. Uniaxial compression tests were conducted along the z-axis to evaluate the effective stiffness of the fabricated lattices. For each configuration, four samples were prepared and tested under the same conditions. The detailed fabrication procedure and experimental process are described in *Materials and Methods*. The resulting stress-strain curves are shown in **Figure 4(E)**, where solid lines represent the average of all tests and shaded areas indicate the range between the minimum and maximum measured values. To ensure consistent comparison, all values were normalized by their respective material stiffness to yield the relative modulus. This eliminates the effect of the base material and highlights the essential role of the structural geometry in mechanical behavior. **Figure 4(F)** presents a comparison of the target property, ML prediction, and experiment results for each case, in relative modulus form. The results show a consistent trend of increasing stiffness across $G_1$ to $G_{10}$, with the experimental values closely matching the intended target values and ML predictions. This agreement demonstrates the effectiveness of the inverse design process using the proposed model. Although a relatively large discrepancy was observed in high stiffness cases compared to the low stiffness cases, we attribute this to geometric irregularities and random point distributions near the edges of the generated point clouds, which can compromise parameter extraction and influence fabrication accuracy. Nevertheless, these experimental results confirm model's capability to generate physically realizable structures with targeted mechanical properties, thereby demonstrating the practical viability of the proposed inverse design framework.

**Extension to other unit cell types**

In order to generalize the presented method beyond BCC, we applied the same inverse design process to different unit cell structures, namely cubic and octahedron structures, as shown in **Figures 5(A)** and **5(F)**. Although the training dataset changed, the proposed framework followed the same procedure, including training the VAE and regressor model, interpolating latent vectors, and generating new unit cell geometries with target properties. Compared to BCC structure, the cubic and octahedron unit cells present increased geometric complexity, including three separate parameters of strut radius $(r_1, r_2, r_3)$ for each unit cell, shown in **Figure 5(A)** and **5(F)**. Despite this added complexity, the model successfully learned the data distribution and constructed a well-aligned latent space with respect to the property $E_z$, along with the directional path for interpolation in latent space. The resulting latent space and interpolation path are illustrated in **Figure 5(B)** and **5(G)**, demonstrating that the proposed framework remains effective for diverse unit cell types. Following the same approach as with the BCC structure, we defined a set of targeted $E_z$ to generate new latent vectors via interpolation along the constructed path. For each unit cell type, four representative cases were selected. The corresponding point cloud geometries were decoded, and their design parameters were extracted to build 3D models. As a result, physical samples of both unit cells and 5×5×5 lattice structures were fabricated. The detailed design parameters and property predictions for each generated structure are summarized in **Tables 2** and **3**. Both the $E_z$ values predicted by the regressor and those obtained via FEA simulation from the decoded structures, showed strong agreement with the predefined target $E_z$, further validating the reliability and flexibility of the inverse design framework. The resulting point cloud outputs, 3D models and fabricated samples are presented in **Figure 5(C)** and **5(H)**.

To experimentally validate the inverse design results of cubic and octahedron structures, fabricated lattice samples were subjected to uniaxial compression tests, following the same procedure as with BCC structure. **Figures 5(D)** and **5(I)** display the resulting stress-strain curves of representative cases for each unit cell, and **Figure 5(E)** and **5(J)** presents the accuracy comparison between target, ML prediction, and experimental values. The experimental values closely followed the trends predicted by the regressor, confirming that the generated structures effectively achieved the target mechanical properties. This agreement further supports the model's ability to generate more complex geometries with multiple design parameters. Similar to the BCC case, the largest errors in the experimental results occurred in samples with high stiffness, with deviations of 17.8% for cubic and 9.4% for octahedron structures. These discrepancies are likely due to two main factors: incomplete hyperparameter optimization during training, and limited availability of training data with high $E_z$, which may have led to less accurate geometry reconstruction in those regions. However, despite the increased complexity of data, the model reliably produced structurally valid and mechanically accurate designs across diverse unit cell types, demonstrating the versatility and robustness of the proposed inverse design framework.

**Design of hybrid unit cells based on multiple unit cell types**

We further extend our approach by training the model with multiple unit cell types simultaneously, in order to construct a unified ML model that accommodates a wider range of geometries. Unlike previous sections, where the model was trained on a single unit cell type, we incorporated six different unit cell types (BCC, Cubic, Kelvin foam, Octahedron, Octet-truss, and Fluorite) into the training dataset. Although this setup poses greater challenges due to increased geometric diversity, it also enables the model to learn structural representations that are more versatile and comprehensive. The training procedure follows the same framework as former section, with additional hyperparameter tuning to account for the heterogeneity of the input data.

**Figure 6(A)** presents PCA visualization of the latent space formed by this multi-structure training. The latent space is clearly divided into six distinct clusters, each representing one of the structural class included in the training dataset. This indicates that the model effectively encodes the structural features of each unit cell type. Moreover, within each cluster, the latent vectors are aligned along the $E_z$ values, demonstrating that the model also captures meaningful property gradients within each structural class. Despite training on various structural types simultaneously, the model effectively reconstructs each unit cell design, closely resembling the original data. These results demonstrate the model's capability to preserve both structural identity and functional properties during the reconstruction process, even in a diverse training environment. However, while property gradients are observable within each cluster, global trend of overall latent space outside the cluster region is not evident. This indicates that while generating new latent vector corresponding to targeted $E_z$ is feasible within individual clusters, understanding the transitions between different unit cell clusters remains challenging.

To investigate the overall latent space, we generated new data within the individual clusters and in the transitional regions between different clusters. First, we generated new data within each cluster, by following the same inverse design process with latent interpolation used in previous sections. A directional path was constructed within each cluster, and new latent vectors were interpolated along this path based on target $E_z$. The results for each cluster are shown under the corresponding unit cell notation in **Figure 6(B)**, with the output structure from each interpolated latent vector. Every resulting structure matches well with the unit cell types of their respective clusters, showing reconstruction accuracy of the VAE model. Next, to explore the latent space beyond individual clusters, we generated new data in the transitional regions between different clusters. Unlike the individual cluster case, constructing a directional path for interpolation between clusters posed a challenge. This difficulty is due to the fact that the train data in the latent space only locate within each cluster, leaving no available reference data points to create path in the inter-cluster regions. Therefore, instead of constructing directional path, we identified data points with similar $E_z$ values from neighboring clusters and connected them to perform linear interpolation between clusters. Rather than depicting a progression from low to high $E_z$ as in previous sections, connecting latent vectors with the identical $E_z$ increases the likelihood of generating latent vector having same $E_z$ value. This approach improves the accuracy of the generated structures and overcomes the difficulty caused by the lack of reference points in the inter-cluster regions. **Figure 6(C)** shows the results of this process, which was repeated for all neighboring clusters and form loop-shaped path that passes through the entire latent space. Linear interpolation was then applied along these loops to generate new latent vectors. Their $E_z$ values were predicted using the regressor, and the corresponding point cloud data were generated via the decoder.

**Figure 6(D)** presents the generation results between different unit cell types. As an example, we generated structures with target $E_z$ of 0.05 GPa. The labels A to F represent the six unit cell configurations positioned along the loop in **Figure 6(C)**. The generated structures are shown next to their respective labels, and corresponding $E_z$ values predicted by the regressor are displayed as a purple line in the central angular chart. The label with two characters, such as A-B, represent the results of interpolation between different unit cell clusters. Remarkably, the model generated hybrid unit cells whose topologies resemble those of the parent unit cells, producing novel structures that were not present in the training dataset. For most of these hybrid unit cells, the predicted $E_z$ were close to the target values. In a few cases, however, the generated designs exhibited different stiffness, with the most notable cases observed in the hybrid between cubic and Kelvin. Interestingly enough, a unit cell with a similar shape was reported in a previous study, where $E_z$ was also higher than expected (*42*). This similarity suggests that the observed deviation reflects the model's ability to capture inherent structural behavior even in unusual cases, rather than a prediction error.

**Figure 6(E)** provides a detailed view of the gradual shape change of hybrid unit cells designed between different unit cells. The labeled shapes at both ends are reconstructed unit cells from the training dataset, while the others are novel configurations designed by the model, illustrating smooth topological transitions between different unit cell types. These newly generated hybrid structures exhibit overall alterations in their topology, as well as dimensional changes in strut radius. For instance, in the hybrid between BCC (A) and cubic (B), the generated data show a gradual transformation where the node at the center of BCC opens, while the diagonal struts split and rotate, and eventually morph into a cubic unit cell. This result demonstrates that the model effectively learns structural features and that the designed paths between clusters in latent space successfully explore them via interpolation, thereby enabling smooth morphological transitions in unit cell designs. Consequently, our approach expands the accessible design space and offers insights into structural evolution between unit cell types.

To validate the properties of the hybrid unit cell designs, we fabricated samples for mechanical testing via 3D printing. Since the hybrid geometries produced by inter-cluster interpolation are too complex for design parameter extraction, as done in previous sections, it is necessary to directly convert the point cloud data into a printable 3D model. We applied the ball-pivoting algorithm for this conversion, preceded by voxel-based point redistribution and a moving-average filtering step to achieve uniform point spacing and preserve fine details. As shown in **Figure 7(A)**, this process successfully converted intricate structures into 3D computer models suitable for 3D printing. Detailed descriptions of the procedure are provided in **Supplementary Materials**. The hybrid unit cells and 5×5×5 lattice samples were fabricated using converted 3D models, as shown in **Figure 7(B)**. The fabricated samples were subjected to uniaxial compression tests to determine their relative Young's modulus. Following the procedure described earlier, the mechanical test results are presented in **Figures 7(C)**, and a comparison between the regressor predictions and the experimental results is provided in **Figure 7(D)**. Unlike previous experiments where multiple target values were used, all samples were designed to achieve a same $E_z$ of 0.02 GPa. The results show that experimental values closely matched the target in most cases. As shown in the property prediction in **Figure 6(D)**, the hybrid unit cell from cubic–Kelvin combination represents an exceptional case with inherently high stiffness. Overall, the average error of all other cases is approximately 16.7%. Considering the complexity of the geometries and potential inaccuracies introduced during the 3D model conversion process, this error is relatively small, experimentally validating the high predictive accuracy achieved through inter-cluster interpolation.

Based on the experimental results, we generated smooth, topologically graded structures connecting two different unit cells via inter-cluster interpolation, while maintaining similar mechanical properties. All generated structures were connected in a single row, showing a smooth unit cell transition process. The fabricated transition structure between BCC and cubic unit cell is shown in **Figure 7(E)**. Other structures showing the transition between different unit cell pairs are provided in **Figure S3** in **Supplementary Materials**. The resulting structures exhibit gradual transitions that naturally bridge the neighboring unit cells, providing intuitive visual evidence that our proposed inverse design method can also generate transitional structures connecting disparate designs. Overall, we successfully demonstrated that the proposed multi–unit cell training approach enables inverse design of intermediate structure between different unit cell types, while reliably achieving the desired mechanical properties. This capability opens a new pathway for creating previously unattainable architectures, expanding the design space for future metamaterial innovations.

**Discussion**

This study presents a rapid ML-driven inverse design method for metamaterials with diverse geometries tailored to specific mechanical properties, addressing the challenges of traditional design approaches such as high computational cost and difficulty to handle complex shape. By integrating deep generative models, specifically a VAE combined with a property regressor, the proposed framework facilitates efficient design and rapid generation of microlattice structures. The model was trained on a dataset of diverse 3D point cloud data paired with corresponding elastic modulus values, allowing for the inverse design of new unit cell geometries with specific mechanical properties. Compared to other 3D representation methods, the use of point cloud data provides significant advantages by enabling the representation of complex, nonparametric structures with high resolution. This suggests that the proposed generative framework can design diverse metamaterial classes within a single unified framework. Additionally, the inclusion of regressor further enhances the framework by eliminating the need for the computationally intensive FEA and enabling property-wise alignment of the latent space. By latent vector interpolation in this aligned latent space, we were able to generate various unit cell geometries with targeted properties. Furthermore, by simultaneous training the model on diverse unit cell classes, we also demonstrated the design of unprecedented hybrid structures that smoothly transform between different unit cell types. This capability allows for the design of topologically graded architectures in which unit cell topologies evolve from one unit cell type to another. Such an approach significantly expands the design space beyond conventional, parameterized methods and opens new possibilities to design innovative metamaterials structures. Finally, the generated point cloud structures were physically fabricated through 3D printing. Experimental validation confirmed that the 3D printed samples exhibited mechanical properties closely matching the predicted values, demonstrating the practical feasibility of the proposed inverse design method.

In this study, we trained the model using six well-established unit cell types, but the training library can be readily expanded to include a broader variety of geometries. While we focused on symmetric unit cells by employing 1/8 segments for computational efficiency, extending the dataset to full unit cells would unlock the design of asymmetric structures with broader functional capabilities. More importantly, the proposed approach can also be extended beyond strut-based lattices to other classes of metamaterials, including shell- and surface-based architectures. With sufficient computational resources, it can be scaled to much larger datasets, providing a versatile and unified framework for data-driven metamaterial design.

**Materials and Methods**

**Machine learning model architecture and training**

The VAE architecture used in this study consists of an encoder with five 1D convolutional layers, efficiently compressing the input data into a meaningful latent space. On the other hand, the decoder is simpler with two layers to prevent overfitting. In both the encoder and decoder, each layer employs the ReLU activation function, enabling the model to effectively learn non-linear patterns in the data (*43*). Additionally, batch normalization is also used in the encoder for fast and stable learning (*44*). The overall architecture of the model is presented in **Figure S2(A)** in **Supplementary Materials**. Point cloud data is inherently permutation invariant, meaning that reordering the input points does not alter its representation. Therefore, to ensure that the output remains consistent regardless of the input point order, we incorporate a max-pooling operation at the end of the encoder, which aggregates features in a permutation-invariant manner. Subsequently, two fully connected neural networks are employed to obtain the mean and variance of the latent vector, for reparameterization trick to sample the latent variable **z** from the obtained mean and variance. After sampling, the latent vector **z** is passed through the decoder, which reconstructs the output data to match the original input's point count. The dataset used for training consists of 10,000 samples per lattice type totaling 60,000 data. These are divided into training, testing, and validation sets in an 8:1:1 ratio. Simultaneously with the VAE, the regressor is also trained to predict the $E_z$ values from the latent vectors of each data. The regressor consists of four fully connected layers, and only uses the mean value of latent vector given by the encoder as its input. The VAE is trained with a learning rate of $10^{-6}$, while the regressor is trained with a learning rate of $10^{-5}$. Training typically involves 500 epochs by default, but early stopping technique was applied to prevent overfitting (*45*). The decrease in loss during the training process and the corresponding early stopping can be observed in **Figure S2(D)** in **Supplementary Materials**. All model training and testing were conducted on a workstation equipped with an NVIDIA GeForce RTX 3090 GPU and 256GB of RAM.

**3D printing of metamaterials**

ML generated metamaterials were fabricated by 3D printing for validation. The lattice structures for mechanical testing were formed by stacking the generated unit cells into 5×5×5 configurations. The test samples were fabricated using a commercial stereolithography (SLA) 3D printer (Form 3L, Formlabs) with a commercially available resin (Clear Resin V4, Formlabs). The mechanical properties of the resin were independently verified through separate testing, which yielded Young's modulus of 1.6 GPa. The 3D printing was conducted with a layer thickness of 100 μm under predefined curing conditions. After printing, the structures were washed in isopropyl alcohol and then fully cured under UV light using post-processing equipment (Cure L, Formlabs) The curing process was performed at 60 °C for 40 minutes to ensure complete polymerization.

**Mechanical testing**

3D printed samples were subjected to compression tests using a mechanical testing system (Criterion Series 40 C43.304 model, MTS). Compressive loads were applied along the vertical direction at a rate of 5 mm/min within a strain range of 4~6%. Each sample was tested more than five times to ensure reproducibility. Young's modulus was calculated from the slope of the elastic region in the stress-strain curve using a linear regression method. Further details are provided in Figure S4 in **Supplementary Materials**.

**Figures and Tables**

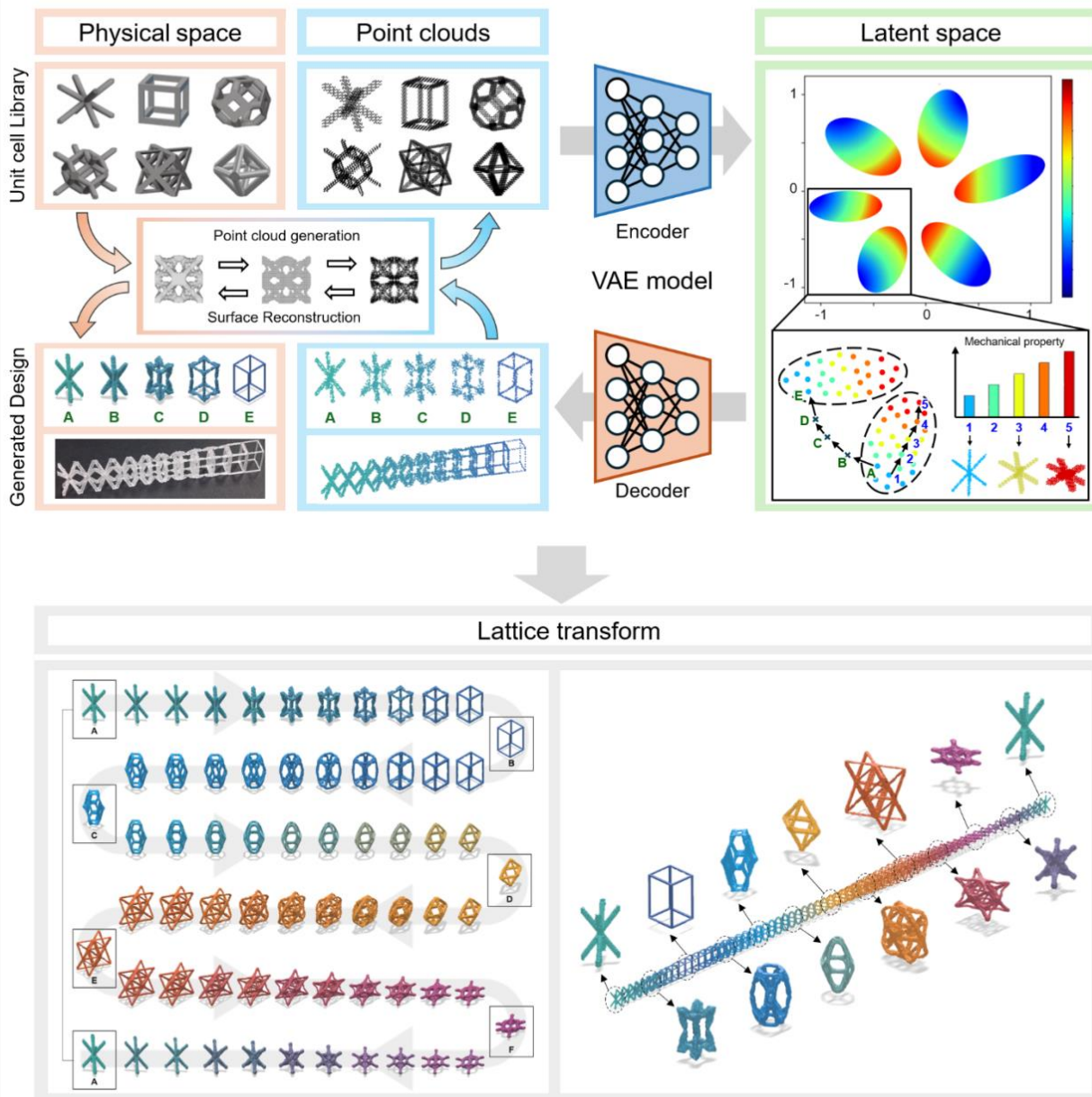

**Figure 1. Overview of machine learning-based generation and analysis of microlattice structures.**

(A)

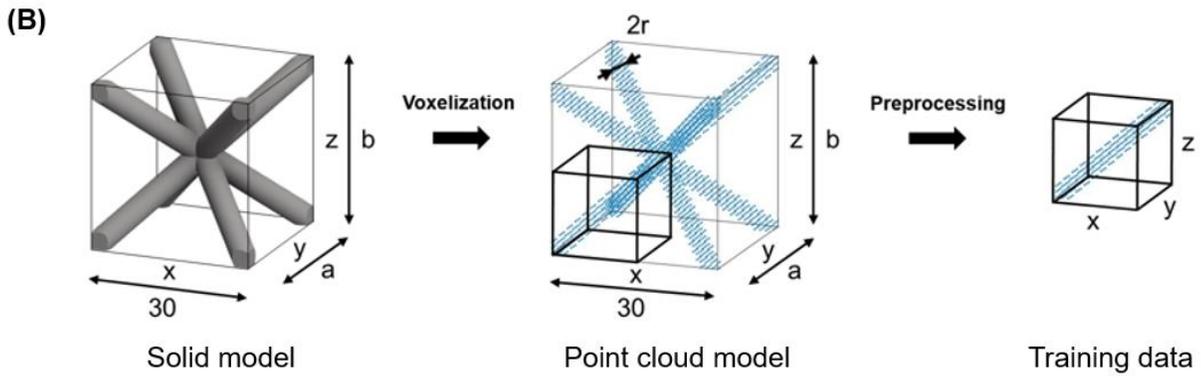

(B)

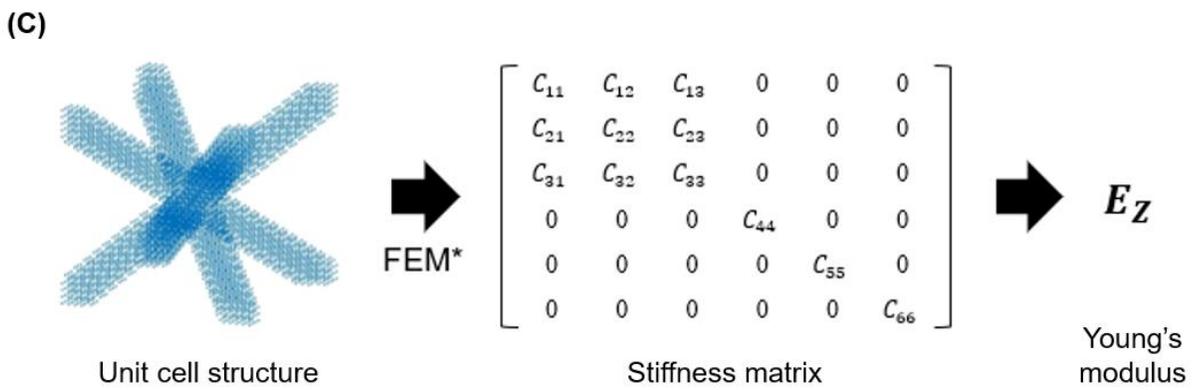

(C)

**Figure 2. Different types of microlattice structures and training data construction.** (**A**) Solid model and point cloud data of six different unit cells (**B**) The generation process and preprocessing of training data. Random values are assigned to *a, b,* and *r* to generate body-centered cubic (BCC) data and followed by the preprocessing step to create training data. (**C**) Property calculation with constructed point cloud data

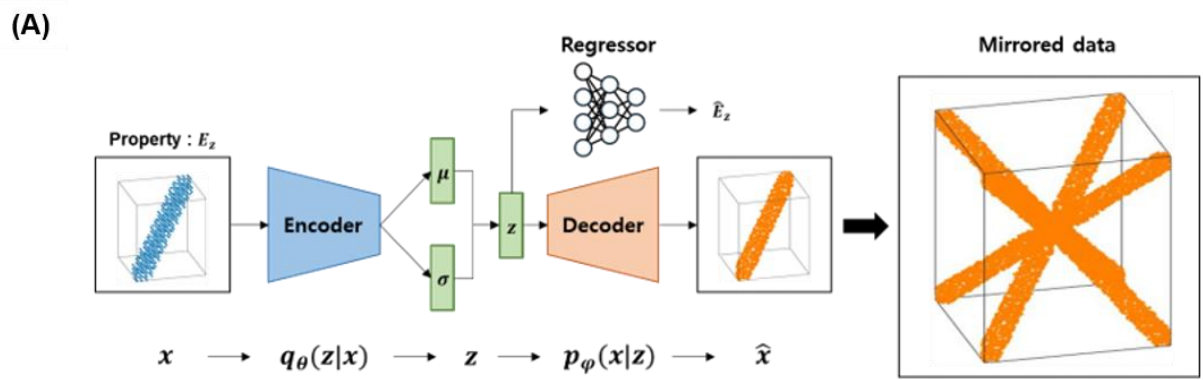

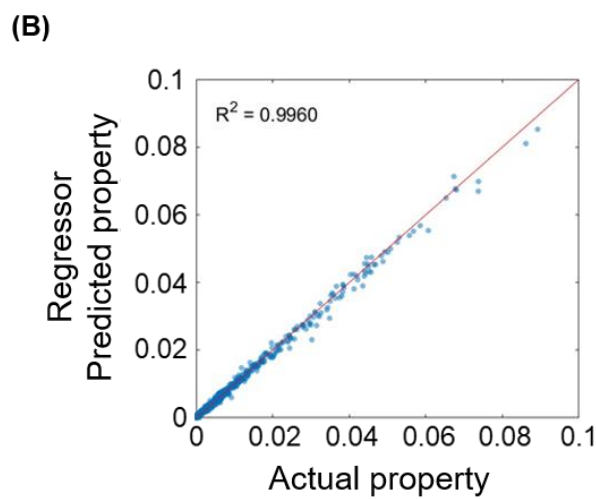
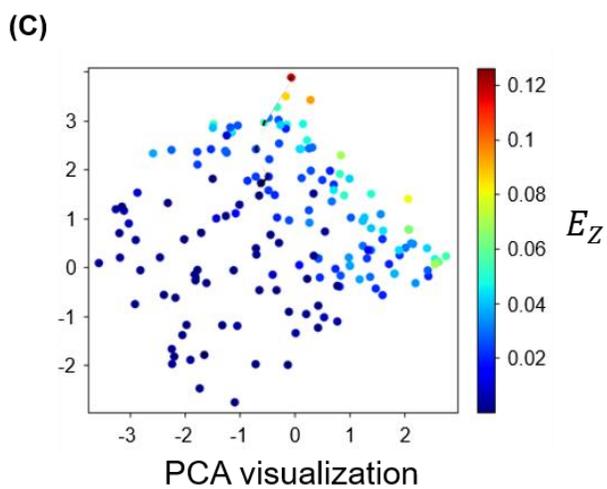

**Figure 3. Machine learning model architecture and results.** (**A**) The schematic architecture of the VAE, along with the mirrored data derived from the output point cloud. (B) Comparison between actual property and predicted result from regressor model. (**C**) Property-aware latent space visualized via PCA

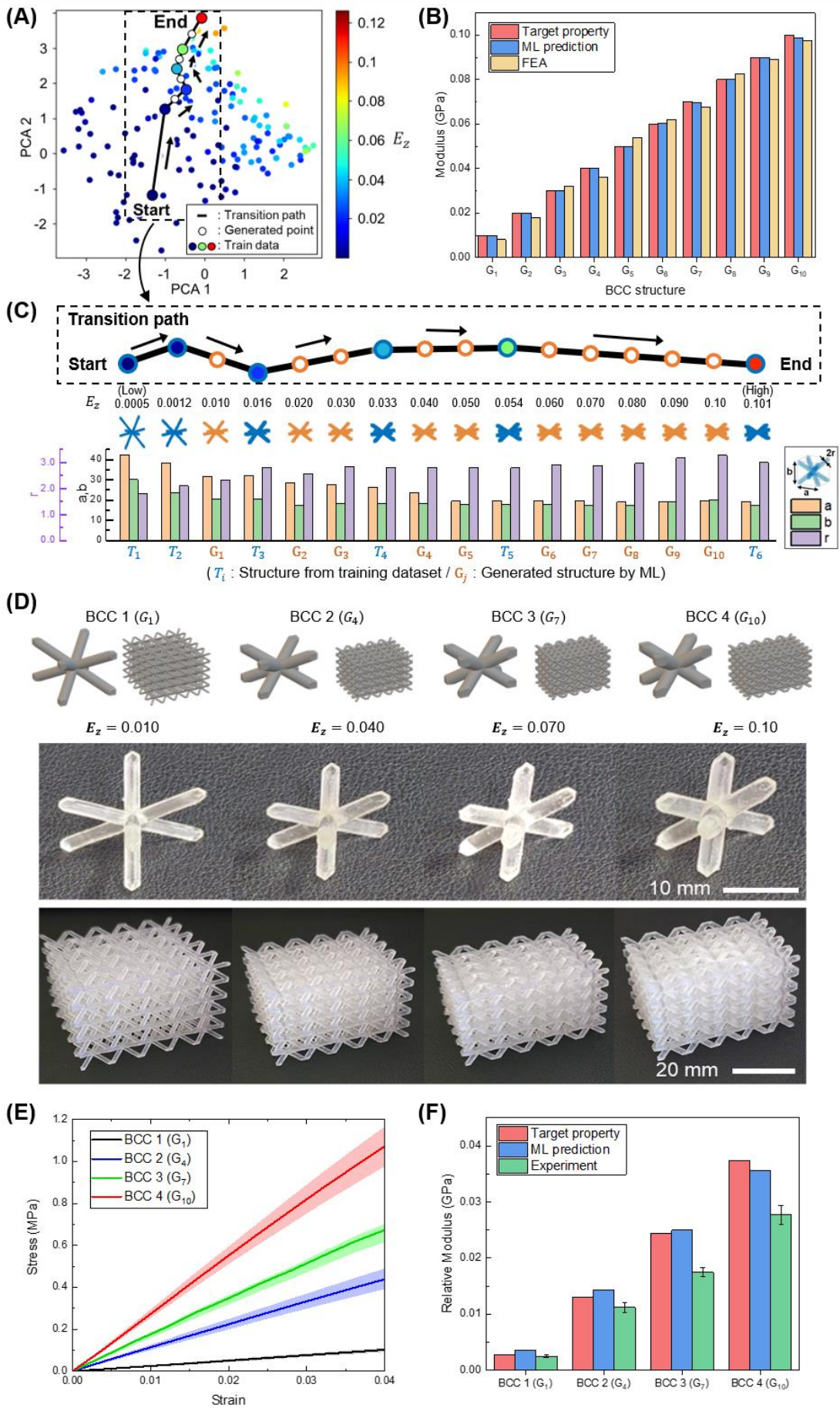

**Figure 4. The generation process for new BCC structures with prescribed $E_z$ values.** (**A**) The directed graph constructed in the latent space. The existing data with colored markers and the new latent vectors generated through linear interpolation with white markers. (**B**) The $E_z$ values of the newly generated data calculated through the ML model and FEA. (**C**) The blue reference structure from the train data and the orange newly generated structure from the interpolated latent vectors, and the variation of the three design parameters in each structure. (**D**) The 3D models of unit cells created from the three data parameters calculated from the newly generated BCC point cloud data, and the lattice structures formed by stacking these unit cells in 5 × 5 × 5 configurations, including those fabricated through 3D printing. (**E**) The stress-strain curves of each lattice obtained from compression tests. (**F**) Comparison between target property, ML prediction, and experimental results for relative modulus.

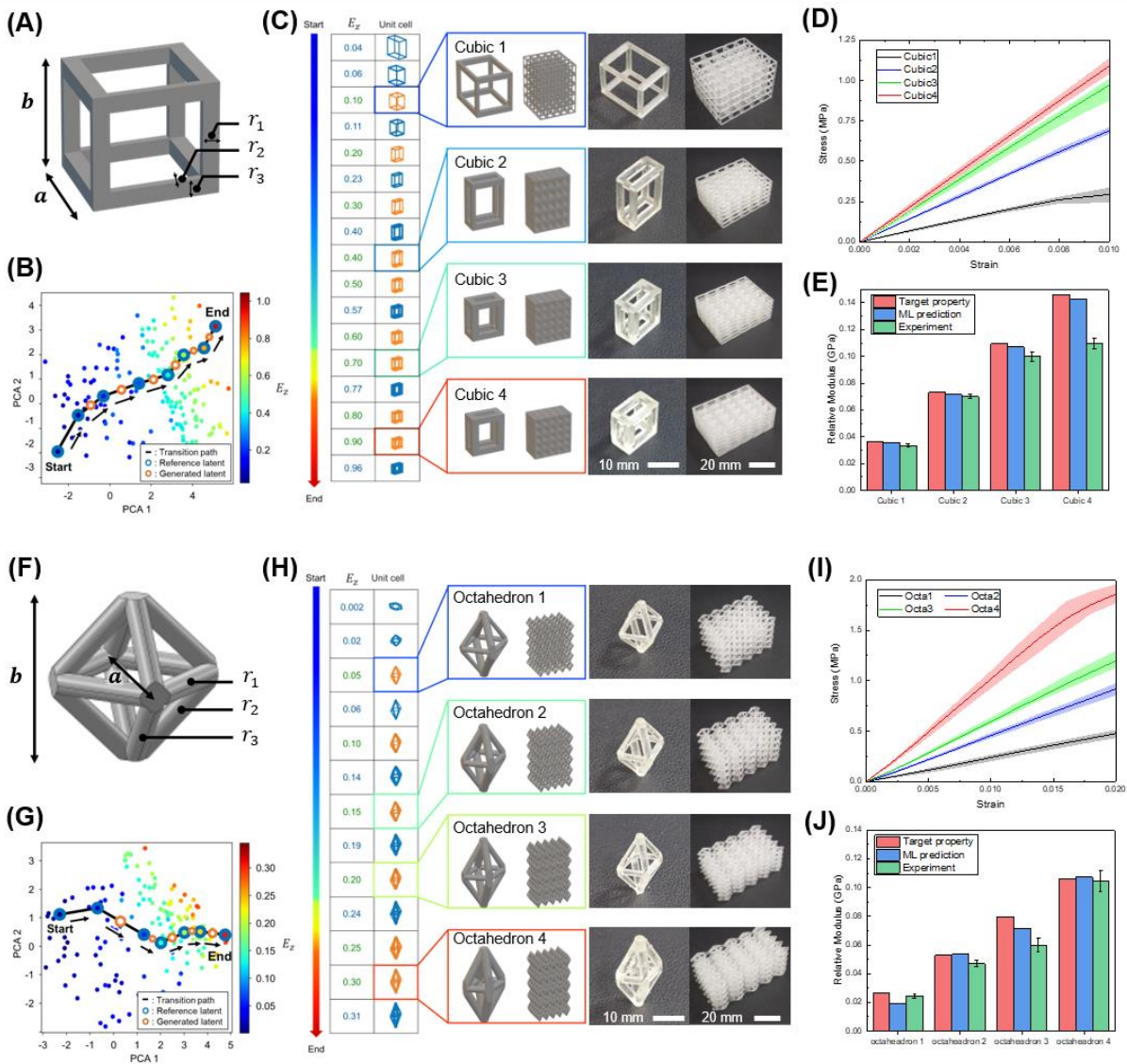

**Figure 5. The generation process for new structures with prescribed $E_z$ values, following the methodology applied to the BCC structures:** (**A-E**) Cubic (**F-J**) Octahedron. (**A**) and (**F**) Design parameters of each structure. (**B**) and (**G**) Latent space and constructed directional path. (**C**) and (**H**) Results of inverse design process and fabricated samples from representative structures. (**D**) and (**I**) Resulting stress-strain curve of compression test. (**E**) and (**J**) Comparison of relative modulus value, between target property, ML prediction and experimental results

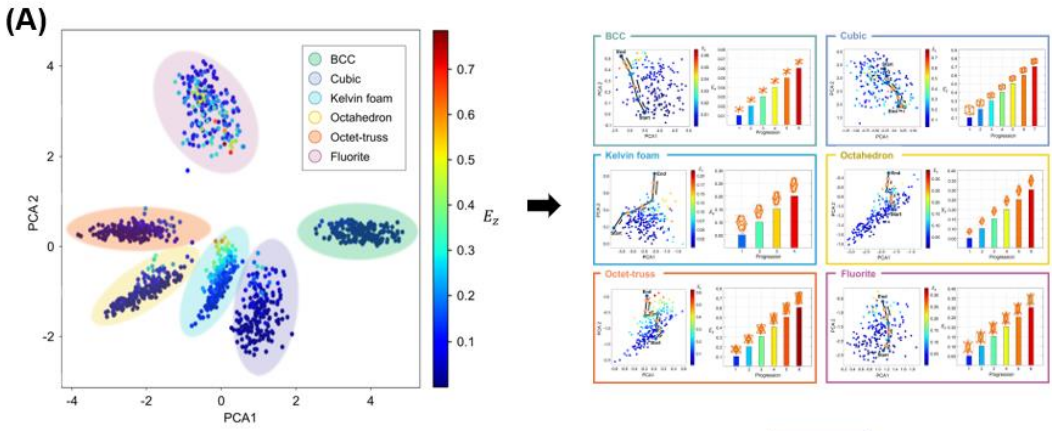

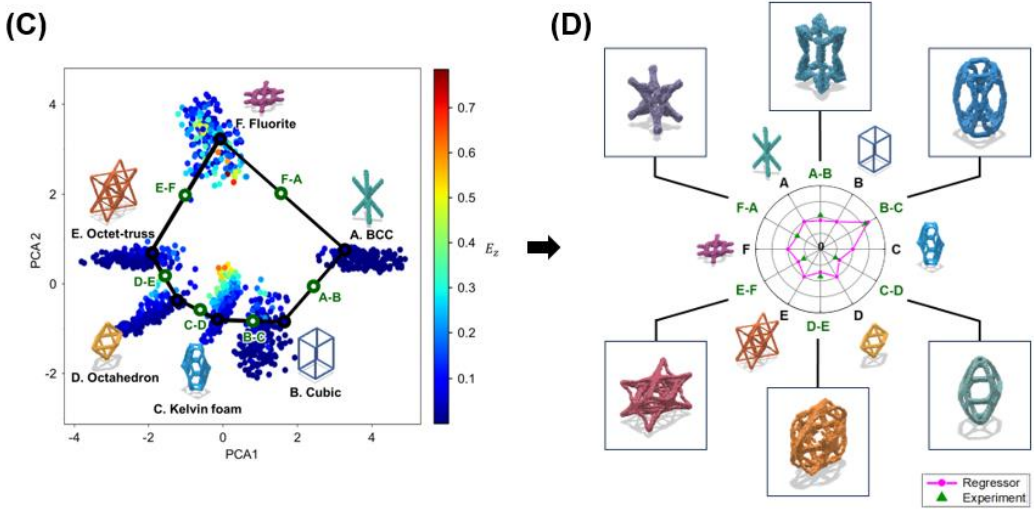

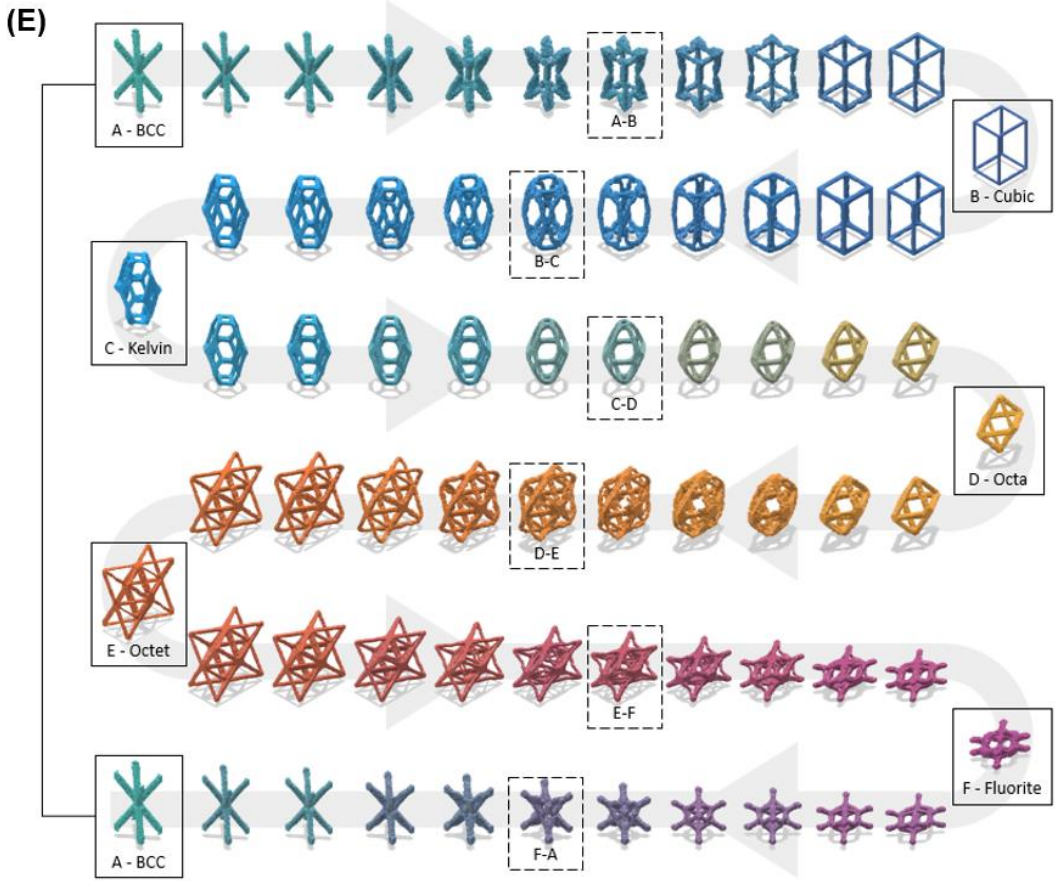

**Figure 6. The results of simultaneous training of diverse structural types.** (**A**) Latent vectors clustered according to unit cell classes, visualized using PCA. (**B**) Inverse design process within each cluster, and newly generated structures with prescribed $E_z$ values are presented. (**C**) Data points with $E_z$ values of 0.050 and their transition path between adjacent clusters to form loops. Green marks indicate newly generated data points within inter cluster region. (**D**) Radar chart showing changes in $E_z$ values for existing unit cell structures and newly generated structures between different unit cell classes. (**E**) The transition process between selected data points, with 10 steps of newly generated interpolation data between each adjacent unit cell class.

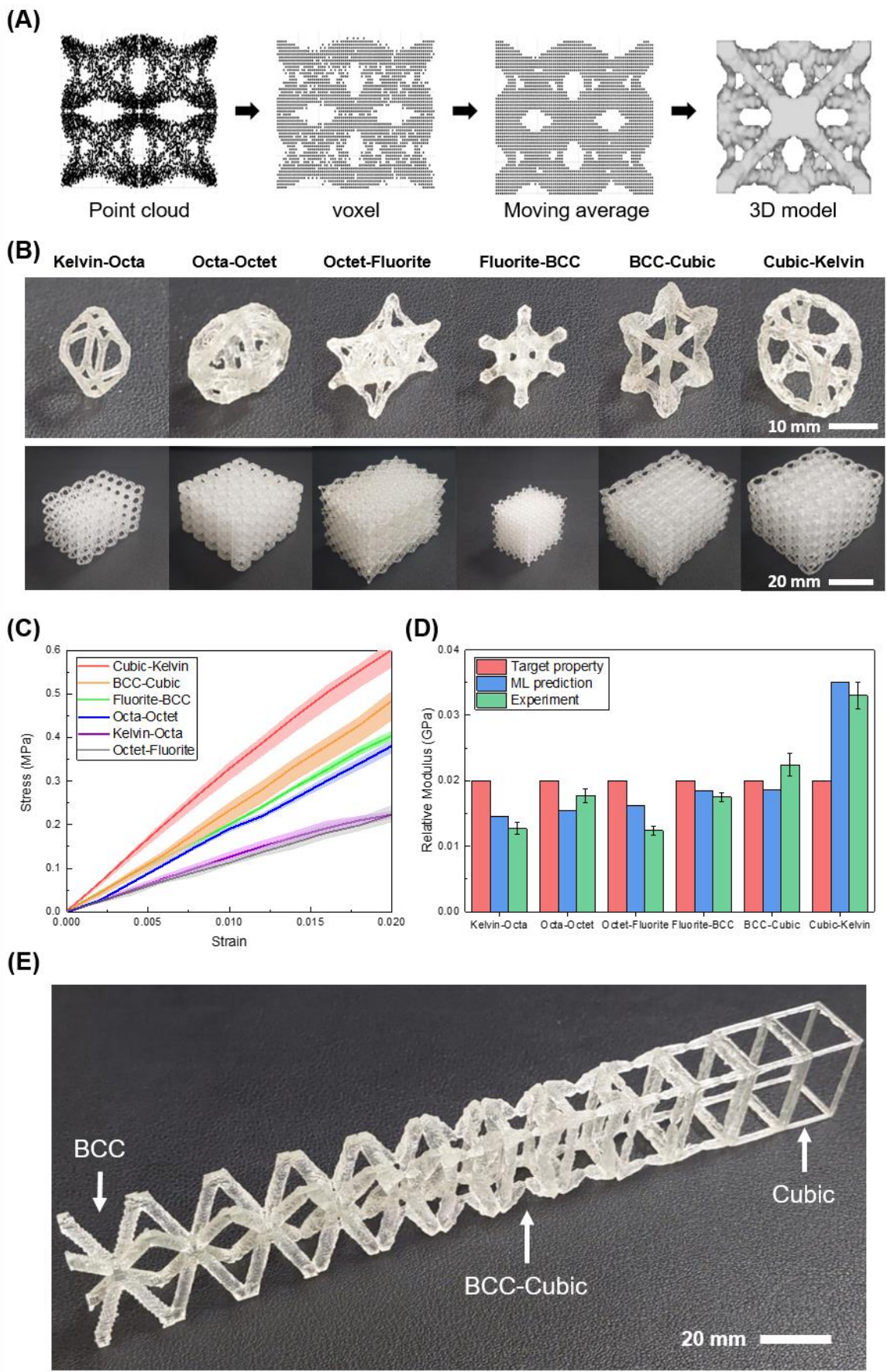

**Figure 7. Fabrication and mechanical testing of lattice structures with interpolated unit cells. (A)**

Surface reconstruction of newly generated point cloud data from the interpolated structure between octet-truss and fluorite lattice, including evenly spaced point clouds, adjustment using the moving average method, and the final surface reconstruction. **(B)** Unit cell and lattice structures fabricated through 3D printing. **(C)** Stress-strain curves obtained from compression tests of each lattice structure. **(D)** Comparison of relative modulus value, between target property, ML prediction and experimental results **(E)** A lattice structure presenting the transition between BCC and cubic unit cells, fabricated as a single connected line using 3D printing.

**Table 1. The three data parameters width (*a*), height (*b*), strut radius (*r*) and $E_z$ calculated from the newly generated BCC point cloud data**

| BCC Structure | Width (a) | Height (b) | Strut radius (r) | $E_z$ (target) | $E_z$ (predicted) | $E_z$ (simulated) |
|---|---|---|---|---|---|---|
| $G_1$ | 31.85 | 20.35 | 2.32 | 0.01 | 0.010 | 0.008 |
| $G_2$ | 28.68 | 17.57 | 2.57 | 0.02 | 0.020 | 0.018 |
| $G_3$ | 27.79 | 18.12 | 2.84 | 0.03 | 0.030 | 0.032 |
| $G_4$ | 23.59 | 18.46 | 2.80 | 0.04 | 0.040 | 0.036 |
| $G_5$ | 19.50 | 17.67 | 2.80 | 0.05 | 0.050 | 0.054 |
| $G_6$ | 19.60 | 17.72 | 2.90 | 0.06 | 0.060 | 0.062 |
| $G_7$ | 19.66 | 17.51 | 2.88 | 0.07 | 0.070 | 0.068 |
| $G_8$ | 19.07 | 17.55 | 2.97 | 0.08 | 0.080 | 0.083 |
| $G_9$ | 19.34 | 19.25 | 3.18 | 0.09 | 0.090 | 0.089 |
| $G_{10}$ | 19.68 | 19.83 | 3.29 | 0.10 | 0.10 | 0.098 |

**Table 2. The five data parameters width (*a*), height (*b*), strut radii ($r_1, r_2, r_3$) and $E_z$ calculated from the cubic point cloud data**

| Cubic Structure | Width (a) | Height (b) | $r_1$ | $r_2$ | $r_3$ | $E_z$ (target) | $E_z$ (predicted) | $E_z$ (simulated) |
|---|---|---|---|---|---|---|---|---|
| $G_1$ | 32.11 | 40.40 | 4.14 | 2.33 | 4.28 | 0.1 | 0.099 | 0.102 |
| $G_2$ | 19.21 | 38.18 | 2.90 | 3.15 | 4.19 | 0.2 | 0.200 | 0193 |
| $G_3$ | 16.77 | 39.96 | 4.12 | 2.92 | 4.72 | 0.3 | 0.299 | 0.294 |
| $G_4$ | 17.26 | 43.69 | 5.24 | 3.16 | 5.29 | 0.4 | 0.402 | 0.362 |
| $G_5$ | 18.13 | 39.02 | 5.45 | 3.78 | 5.24 | 0.5 | 0.500 | 0.456 |
| $G_6$ | 18.78 | 35.91 | 5.38 | 4.48 | 5.28 | 0.6 | 0.600 | 0.644 |
| $G_7$ | 17.99 | 34.26 | 5.30 | 5.83 | 5.20 | 0.7 | 0.699 | 0.683 |
| $G_8$ | 17.89 | 32.87 | 5.60 | 5.65 | 5.25 | 0.8 | 0.800 | 0.816 |
| $G_9$ | 16.34 | 31.61 | 5.76 | 5.99 | 5.30 | 0.9 | 0.901 | 0.917 |

Table 3. The five data parameters width ($a$), height ($b$), strut radii ($r_1, r_2, r_3$) and $E_z$ calculated from the octahedron point cloud data

| Octahedron Structure | Width ($a$) | Height ($b$) | $r_1$ | $r_2$ | $r_3$ | $E_z$ (target) | $E_z$ (predicted) | $E_z$ (simulated) |
|---|---|---|---|---|---|---|---|---|
| $G_1$ | 23.85 | 40.33 | 2.36 | 1.64 | 1.52 | 0.05 | 0.050 | 0.078 |
| $G_2$ | 24.60 | 44.68 | 2.51 | 1.74 | 1.88 | 0.10 | 0.100 | 0.104 |
| $G_3$ | 24.45 | 46.05 | 2.67 | 2.00 | 2.15 | 0.15 | 0.150 | 0.143 |
| $G_4$ | 17.65 | 47.70 | 2.52 | 1.88 | 2.11 | 0.20 | 0.199 | 0.193 |
| $G_5$ | 19.75 | 52.10 | 2.96 | 2.19 | 2.57 | 0.25 | 0.250 | 0.265 |
| $G_6$ | 20.43 | 54.19 | 3.18 | 2.47 | 3.09 | 0.30 | 0.299 | 0.329 |

# Supplementary Materials for

## Point-Cloud Based Inverse Design of Free-Form Metamaterials Using Deep Generative Networks

Kijung Kim[1†], Seungwook Hong[1†], Wonjun Jung[1], Wooseok Kim[1], Namjung Kim[2*], and Howon Lee[1*]

[1]Department of Mechanical Engineering, Institute of Advanced Machines and Design, Seoul National University, Seoul, Republic of Korea

[2]Department of Mechanical Engineering, Gachon University, Sungnam, Republic of Korea

[*]To whom correspondence should be addressed: namjungk@gachon.ac.kr, howon.lee@snu.ac.kr

[†]These authors contributed equally to this work.

**This PDF file includes:**

    Supplementary Text
    Figs. S1 to S4
    Tables S1 to S2
    References (#1)

**Data generation for other unit cell types**

For each unit-cell geometry (BCC, cubic, octahedron, octet-truss, kelvin foam, and fluorite), we generated point cloud data by varying structural parameters such as width, height and strut radius. Each unit cell structure contains a different number of struts, and the radius of each strut can also be parameterized. By assigning random values to the selected parameters, a complete wireframe structure can be created. Point cloud data were then generated by embedding this wireframe in a 3D voxel grid and calculating the distance from each voxel center to the nearest wireframe segment, retaining points within the specified strut radius. Since each dataset has a different structural size, structures were normalized to fit within a unit cube by centering them at the origin and applying uniform scaling. Additionally, to standardize the number of points per structure, a padding process was applied by duplicating points farthest from the origin until the desired number of points was reached. This padding strategy helped preserve the structural integrity and improved the reconstruction stability during training. Finally, to enhance efficiency, the inherent symmetry of each structure was leveraged to compress the data to 1/8 of its original size Through this preprocessing, consistency across all datasets was ensured, and the overall process is illustrated in **Figure S1**.

**Training process**

Optimization of hyperparameters

The VAE architecture used in this study consist of encoder, decoder, and regressor, which illustrated in **Figure S2(A).** To achieve robust training of this model, we optimized hyperparameters to enhance reconstruction quality and effective clustering in latent space, ensuring the VAE captures essential geometric features. In our case, we focused significantly on enhancing reconstruction quality. Specifically, to improve the reconstruction quality of the VAE, we optimized four hyperparameters: the dimensions of the latent space ($D$), and the weights of the regularization loss ($\beta$), the regression loss ($\gamma$) and contrastive learning loss ($\delta$). Sequentially, while keeping the other hyperparameters constant, we adjusted each hyperparameter in turn to find the values that minimized the reconstruction loss which is Chamfer distance between input and output data. Moreover, this process was repeated to determine the optimal values. As a result, as shown in **Figure S2(B)**, the optimal values for the latent dimension and $\gamma$ are 22 and $10^3$, respectively. Regarding $\boldsymbol{\beta}$, we noted that the reconstruction loss tended to decrease as $\boldsymbol{\beta}$ was reduced. However, excessively low values of β could impair the VAE's ability to learn meaningful latent representations and lead to poor generalization in generating new data. As shown in **Figure S2(B)**, there was no significant improvement in reconstruction loss for $\boldsymbol{\beta}$ values below $10^{-4}$. Therefore, to balance reconstruction quality with the VAE's performance, we selected $10^{-4}$ as the optimal value for $\boldsymbol{\beta}$. Regarding $\boldsymbol{\delta}$, since it has more influence on contrastive clustering in the latent space according to property, we adjusted it around the optimal value of $10^{-3}$ to observe how clustering occurs. In **Figure S2(C)**, the PCA (Principal Component Analysis) results are shown for the case when $\boldsymbol{\delta}$ is set to $10^{-3}$ and $10^{-2}$. However, there doesn't seem to be a significant difference between the two results, we ultimately selected $10^{-3}$ as the optimal value for $\boldsymbol{\gamma}$. This choice was based on a combination of the observed clustering patterns in the PCA analysis and the overall goal of achieving a balance between accurate reconstruction and meaningful clustering based on properties in the latent space.

Training results

After optimization process, the training results, as shown in **Figure S2(D)**, showed that both the training loss and validation loss decreased steadily over the epochs. Training concluded near 400 epochs due to early stopping and the regressor was trained similarly. Using the encoder, we obtained latent vectors from the test data, and then used the regressor to estimate the $E_z$ values from these vectors. Upon comparing these estimates with the actual $E_z$ values, as shown in **Figure S2(E)**, the $R^2$ value was 0.9960, indicating a high degree of agreement with the actual values. Additionally, **Figure S2(F)** shows the reconstruction result of test data using the VAE. Not only does the overall shape of the reconstructed data resemble the original, but the projections of the data onto the xy-plane, yz-plane, and xz-plane also show a high degree of similarity. However, the reconstructed data appears to be distributed more irregularly with noise, unlike the original data arranged in a regular grid. When we calculated the three data parameters $\boldsymbol{a, b, r}$ of the reconstructed data, we observed error rates of 2.55%, 2.73%, and 3.65%, respectively, compared to the original test data.

**Transition unit cell**

By interpolating the latent vectors between two different unit cell structures and reconstructing them through the decoder, a transition unit cell structure connecting the distinct unit cells can be formed. Also, we can adjust the interpolation degree, to create a transition unit cell that is closer to a specific unit cell, or make the transition unit cells changing gradually from one unit cell to others. Since the adjacent transition unit cells on gradual changing line have similar shapes, connectivity can be established between them and allows the transformation process between the two unit cell structures to be represented as a connected 1D line lattice. All six transformations structures are represented in **Figure S3**

**Conversion of point cloud data into 3D model**

To validate the properties of the newly generated structures, we attempted to fabricate them through additive manufacturing. However, this process posed a significant challenge, as the generated point cloud data represented novel geometries not present in the original dataset. Unlike previous cases, where simple parameters such as strut thickness could be extracted and used for validation, the unfamiliar structure of these data made such parameterization infeasible. Therefore, a conversion step was required to transform the point cloud representations into manufacturable 3D models. While point cloud data are useful for structural visualization, additive manufacturing requires surface or volume -based data. Among various methods developed for this purpose, the ball-pivoting algorithm is one of the most widely used approaches for reconstructing watertight surface meshes from point cloud data. (*1*). This method creates surface meshes by *pivoting* a ball around the points to form triangular surfaces. **Figure 6(A)** in the main text demonstrates the conversion process from raw point cloud data to a surface mesh. Due to its operational principle, the ball-pivoting algorithm is most effective when every point in the point cloud is uniformly distributed, which is not the case with our model's randomly distributed point clouds shown in the first steps of **Figure 6(A)**. To address this issue, we divided the point cloud into voxels and repositioned each point to the center of the nearest voxel, thereby achieving an even distribution. While this process improves surface reconstruction, it can also result in the loss or distortion of fine structural details, as seen in the second step. To mitigate this limitation, a moving average method is applied. In this method, each voxel was examined based on the density of surrounding points, and the voxel center was preserved only if the local point density exceeded a set threshold. The resulting evenly distributed point cloud and its surface reconstruction using the ball-pivoting algorithm are shown in third and fourth step of **Figure 6(A)**. This method proved effective even for complex geometries, such as interpolated structures between octet-truss and fluorite Through this process, we successfully converted the novel interpolated structure into 3D surface models, enabling the fabrication of lattice structures.

**Mechanical testing**

From our VAE based model, point cloud data of the target unit cell structure can be generated based on the latent vector, and the same latent vector can be used by the regressor to predict $E_z$ of structure. To validate the accuracy of the entire model, including the regressor, we fabricated the generated structures through 3D printing and conducted experiments. With surface reconstruction process, the resulting point cloud structure was converted to surface data, which is suitable for additive manufacturing. Surface data of unit cell structure were arranged into a $5 \times 5 \times 5$ lattice using 3D CAD software, and lattice samples for mechanical testing were printed. The tests were conducted on three types of unit cell, BCC, cubic, and octahedron. For each unit cell type, four different structures having variations in parameters such as strut radius are decided, such as BCC1, BCC2, BCC3 and BCC4. Then, five samples are printed for each structure to conduct repeated experiment. To minimize the effect of the printing orientation, all lattice models were compressed along the direction perpendicular to the layer interface aligned in the same direction relative to the compression axis, ensuring consistent slicing conditions and accurate mechanical comparison.

The printed samples underwent compression testing using a customized mechanical tester, as shown in **Figure S4(A)**. Each sample was manually loaded and compressed at a rate of 5 mm/min, and the stress-strain curve was measured up to 4~6% strain. The raw experimental data of BCC unit cell structure 4 is presented in **Figure S4(B)**, showing that the initial part of the stress-strain curve is unreliable due to inaccuracies during manual loading, leading to an improper start of the test. To identify appropriate test region, we analyzed the data using linear regression method and found the linear section of stress-strain curve, indicating the accurate compression area. $E_z$

was measured from the slope of this linear section. We applied the same process to repeated experiments for other samples with same structure, and confirmed the linear region of structure, which is combined and illustrated in **Figure S4(C).** The same experiments were conducted on all samples of the BCC unit cell structure, and the results are presented in **Figure S4(D)**, where they are simplified by using mean data line and a shaded area. We repeated this process for the cubic and octahedron structures as well. $E_z$ obtained from the experiments was converted into a relative modulus by dividing by the material's modulus ($E_M$) and all the data are shown in **Table S1**. Table also includes the regressor results from the machine learning (ML) model and the homogenization method, used as the ground truth. Additionally, it shows the absolute and percentage errors for each comparison between the experimental data and the ML model, and between the experimental data and the results from the homogenization method. The experimental results demonstrated an average error of less than 20% from these values, indicating allowable level of accuracy.

For transition unit cells between two different unit cell types, we represented six different types of transition unit cells by mapping positions between adjacent unit cell types in the latent space from ML model. These transition unit cells were tested using the same procedures as the single unit cell types. Since transition unit cells were not included in the training dataset and could not be parameterized into existing known shapes, ground truth values cannot be calculated as single unit cell. Therefore, only the modulus values predicted by the ML model regressor were presented and compared with the experimental results, as shown in **Table S2**.

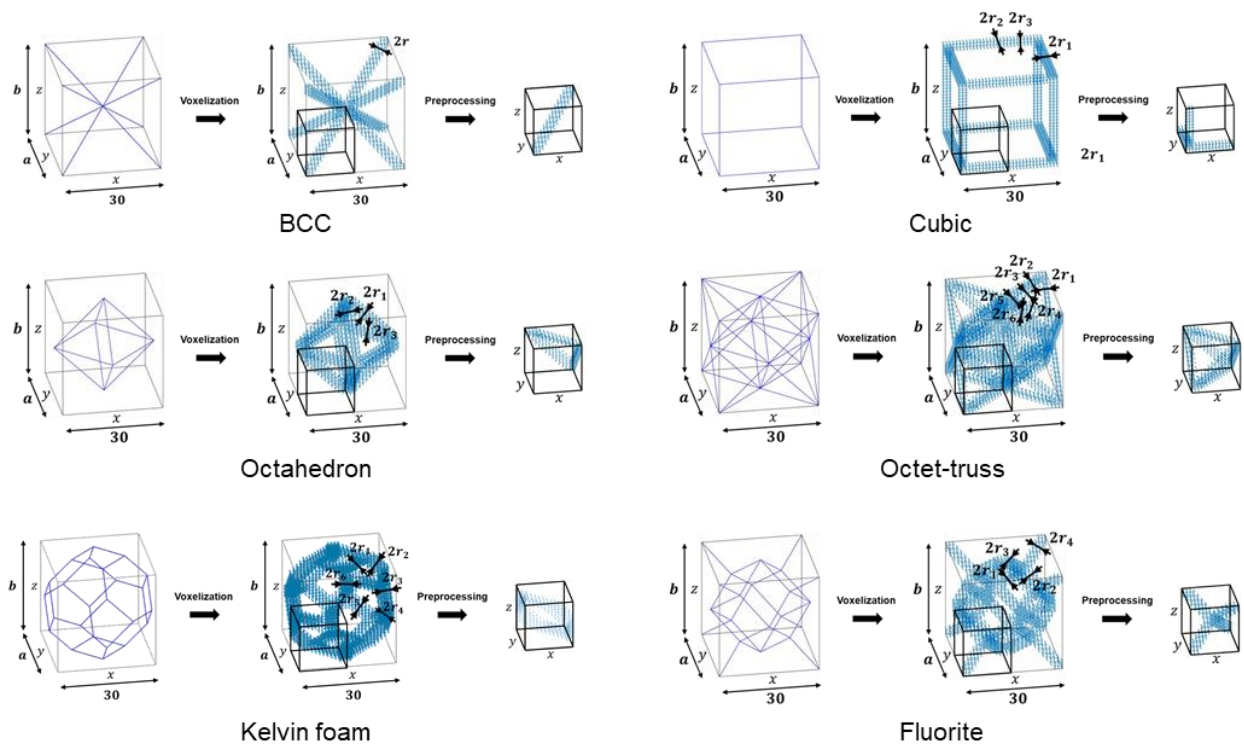

**Figure S1. Point cloud data generation from six-unit cell geometries.** Point cloud generation based on Design parameters (height, width, radius) for each structure, and creation of training dataset by excluding symmetric parts.

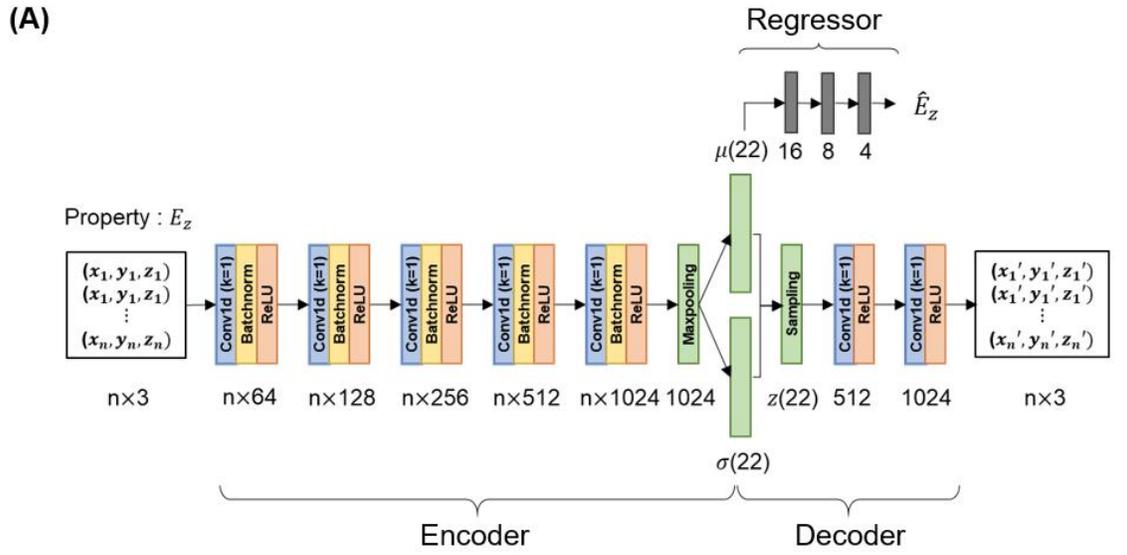

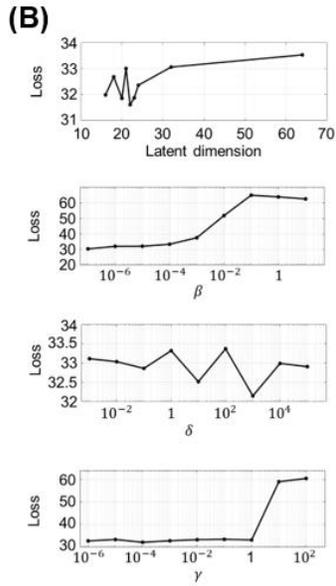
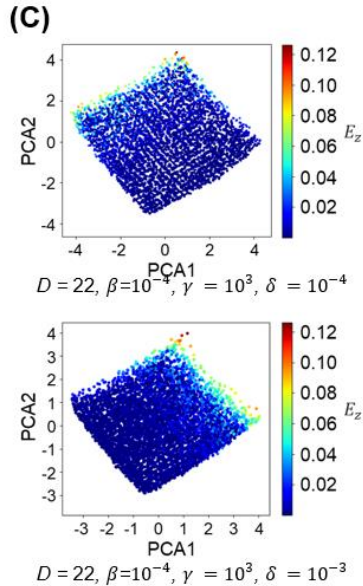
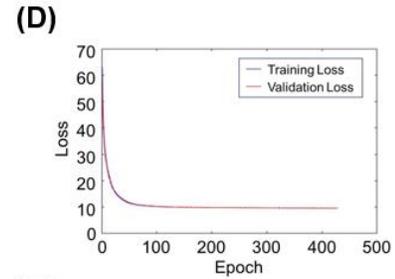
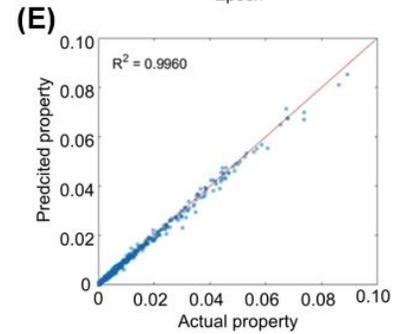

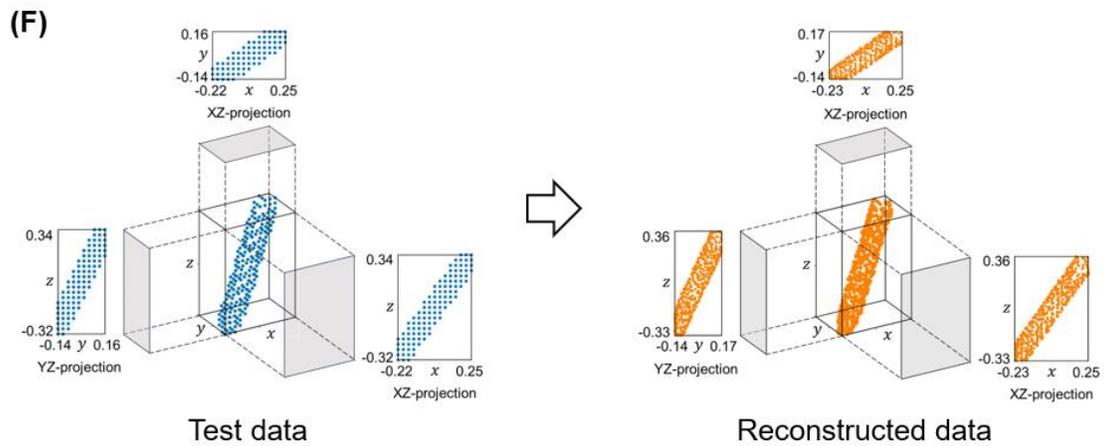

**Figure S2. Machine learning model and the training results. (A) Overall VAE model architecture (B) The results of the optimization for each latent dimension and hyperparameters $\beta, \gamma$, and $\delta$. (C) The different PCA results across varying $\delta$ values $10^{-4}$ and $10^{-3}$.(D) changes in the training loss and validation loss according to the epoch. (E) The comparison of the actual $E_z$ values of test data with the values predicted by the regressor. (F) The point cloud data comparison between the test data and the reconstructed data through the ML model, along with the images projected onto the xy-plane, yz-plane, and zx-plane.**

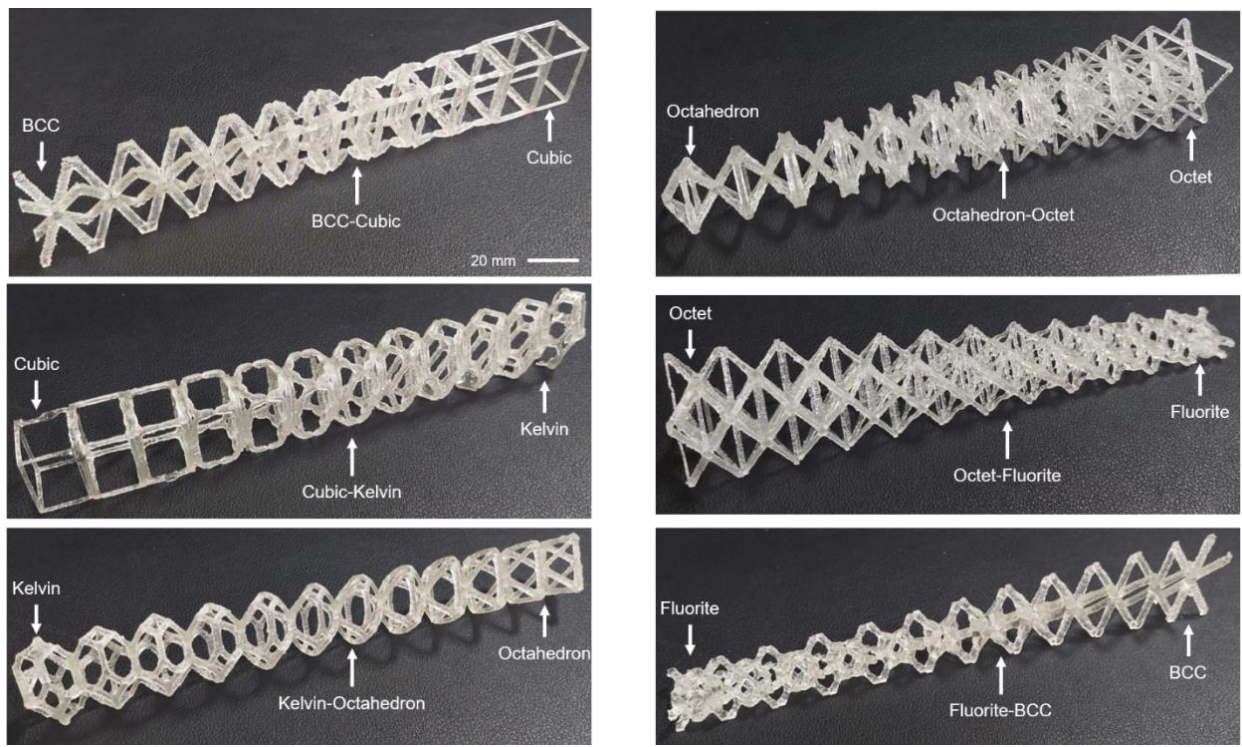

**Figure S3. 3D-printed results of unit cell shape transition.** Linearly connected structure of 2 different unit cells and the interpolated shapes between them, demonstrating smooth transitions between different geometries

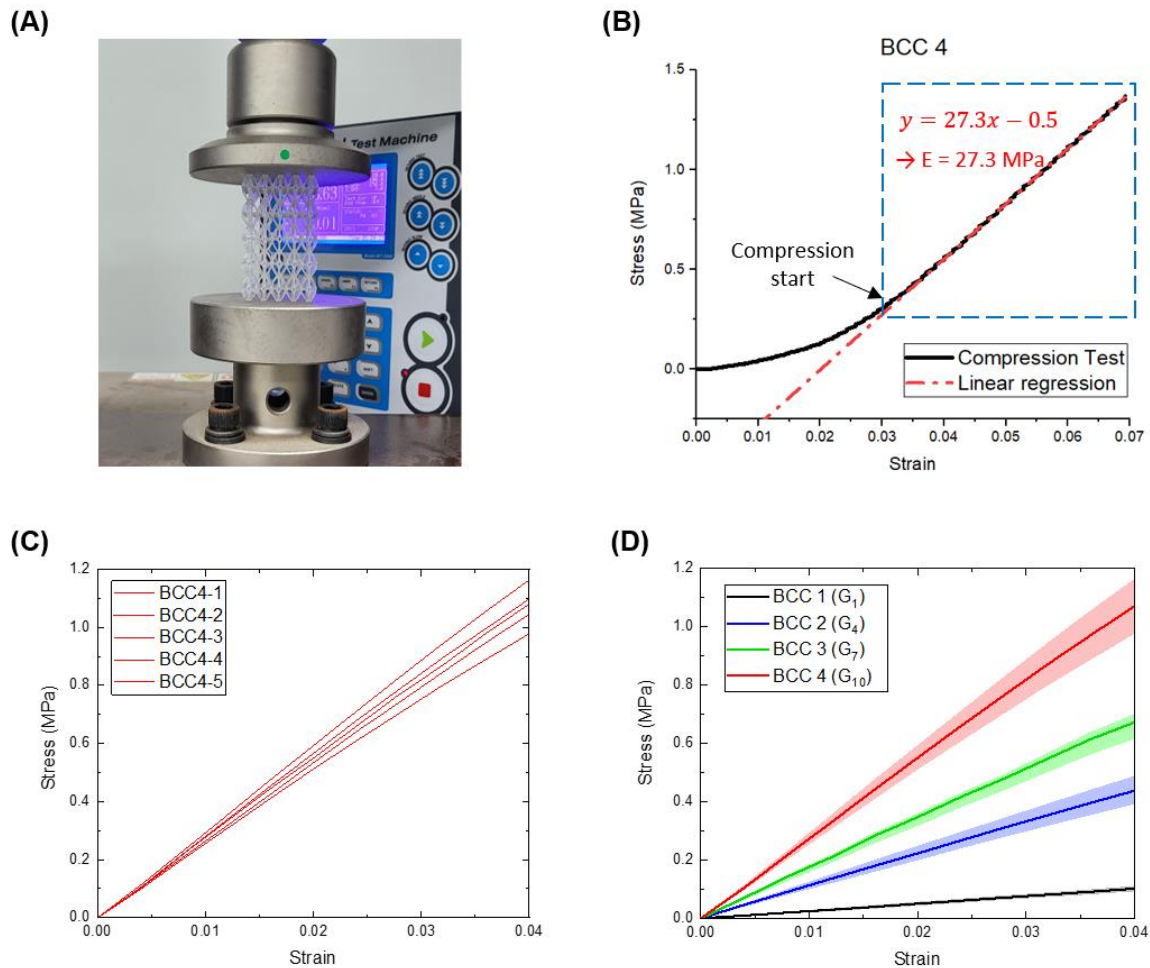

**Figure S4. Lattice compression test results.** (A) Compression tests setup using custom mechanical tester. (B) Raw test data and analysis through linear regression, which allowed identification of the actual compression section and modulus. (C) Plot showing repeated experiments results that represent only the actual compression regions. (D) Combined test results for samples with the same unit cell structure but different design parameters.

|  | Volume fraction | Relative modulus (E'/E, unit: 1e-3) | | | Experiment error | |
| --- | --- | --- | --- | --- | --- | --- |
|  |  | Target property ($E_T$) | ML model ($E_{ML}$) | Experiment ($E_{Exp}$) | $E_T - E_{Exp}$ | $E_{ML} - E_{Exp}$ |
| BCC 1 | 0.17 | 2.8 | 3.5 | 2.5 | 0.3 (10.7%) | 1.0 (28.6%) |
| BCC 2 | 0.32 | 12.9 | 14.3 | 11.2 | 1.7 (13.2%) | 3.1 (21.7%) |
| BCC 3 | 0.40 | 24.4 | 25.0 | 17.5 | 6.9 (28.3%) | 7.5 (30.0%) |
| BCC 4 | 0.48 | 37.4 | 35.7 | 27.8 | 9.6 (25.7%) | 7.9 (22.1%) |
| CUBIC 1 | 0.13 | 36.5 | 35.7 | 33.5 | 3.0 (8.2%) | 2.2 (6.2%) |
| CUBIC 2 | 0.30 | 71.4 | 71.4 | 70.2 | 2.0 (2.8%) | 1.2 (1.7%) |
| CUBIC 3 | 0.53 | 110.5 | 107.1 | 99.8 | 10.7 (9.7%) | 7.3 (6.8%) |
| CUBIC 4 | 0.65 | 129.3 | 142.9 | 110.0 | 19.3 (14.9%) | 32.9 (23.0%) |
| OCTA 1 | 0.11 | 26.9 | 19.1 | 24.5 | 2.4 (8.9%) | -5.4 (-28.3%) |
| OCTA 2 | 0.15 | 54.1 | 53.6 | 47.1 | 7.0 (12.9%) | 6.5 (12.1%) |
| OCTA 3 | 0.18 | 68.8 | 71.4 | 60.0 | 8.8 (12.8%) | 11.4 (16.0%) |
| OCTA 4 | 0.26 | 117.6 | 107.1 | 104.4 | 13.2 (11.2%) | 2.7 (2.5%) |

**Table S1. Compression test results for generated lattice structures based on a single type of unit cell (BCC, Cubic, Octahedron)**

|  | Volume fraction | Relative modulus ($E_L/E_M$, unit: 1e-3) | | Experiment error |
|---|---|---|---|---|
|  |  | ML ($E_{ML}$) | Experiment ($E_{Exp}$) | $E_{ML}$ - $E_{Exp}$ |
| Fluorite - BCC | 0.30 | 16.4 | 16.5 | -0.1 (0.6%) |
| BCC - Cubic | 0.21 | 17.1 | 22.4 | -5.3 (31.0%) |
| Cubic - Kelvin | 0.16 | 35.0 | 33.1 | 1.9 (5.4%) |
| Kelvin - Octa | 0.08 | 11.8 | 10.7 | 1.1 (9.3%) |
| Octa – Octet | 0.15 | 12.5 | 16.2 | -3.7 (29.6%) |
| Octet - Fluorite | 0.21 | 13.7 | 9.4 | 4.3 (31.4%) |

**Table S2. Compression test results for generated lattice structures of transition shape between 2 different types of unit cell**

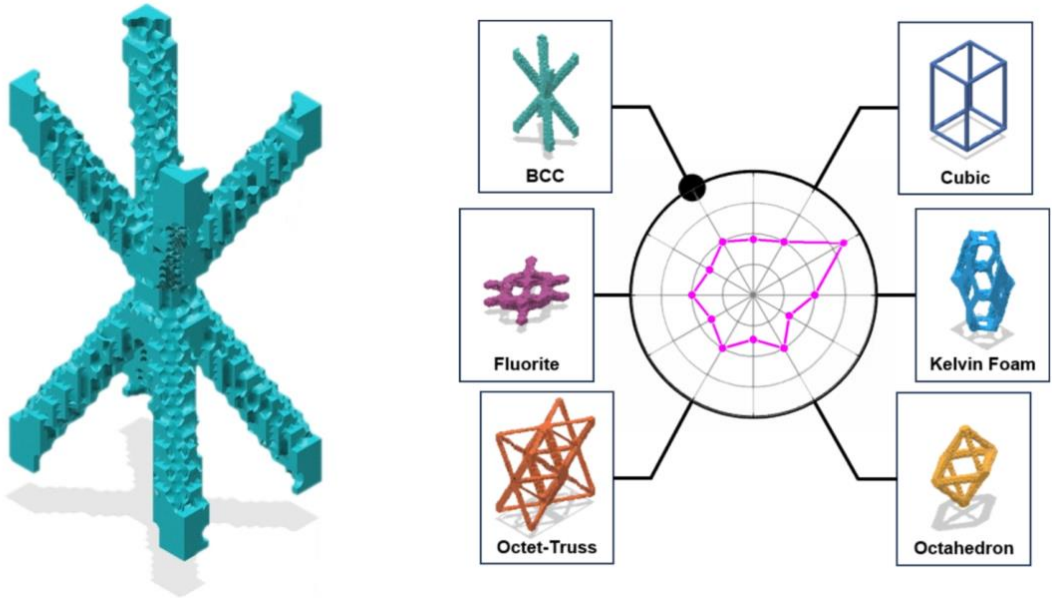

**Movie S1. Transition of 3D Mechanical Metamaterial structures and properties.** An animation showing the shape transitions and respective mechanical properties among six unit cell structures, generated by the machine learning model and surface reconstruction process.